\documentclass[letterpaper, 10 pt, conference]{ieeeconf}
\IEEEoverridecommandlockouts                              

\IEEEoverridecommandlockouts
\usepackage{cite}
\usepackage{amsmath,amssymb,amsfonts}
\usepackage{subfigure}
\usepackage{graphicx}
\usepackage{textcomp}
\usepackage{xcolor}
\usepackage{tabularx}
\usepackage{array}
\usepackage{braket,booktabs}
\usepackage{threeparttable}  
\usepackage{multirow}
\usepackage{siunitx,booktabs} 
\renewcommand{\arraystretch}{1.2}
\newcommand{\ra}[1]{\renewcommand{\arraystretch}{#1}}
\usepackage{bm}
\usepackage{amssymb}
\usepackage{pifont}
\usepackage{algorithm}
\usepackage{algpseudocode}

\def\BibTeX{{\rm B\kern-.05em{\sc i\kern-.025em b}\kern-.08em
    T\kern-.1667em\lower.7ex\hbox{E}\kern-.125emX}}

\begin{document}

\title{\LARGE \bf
Decoding Human Attentive States from Spatial-temporal EEG Patches Using Transformers
}

\author{Yi Ding, Joon Hei Lee, Shuailei Zhang, Tianze Luo,
  Cuntai Guan*
  \thanks{
  Yi Ding, Joon Hei Lee, Shuailei Zhang, Tianze Luo and Cuntai Guan are with the Collage of Computing and Data Science, Nanyang Technological University, Singapore. 50 Nanyang Ave, Singapore 639798.
  * Cuntai Guan is the Corresponding Author. Email: ctguan@ntu.edu.sg}
}
\maketitle

\thispagestyle{empty}
\pagestyle{empty}


\begin{abstract}
Learning the spatial topology of electroencephalogram (EEG) channels and their temporal dynamics is crucial for decoding attention states. This paper introduces EEG-PatchFormer, a transformer-based deep learning framework designed specifically for EEG attention classification in Brain-Computer Interface (BCI) applications. By integrating a Temporal CNN for frequency-based EEG feature extraction, a pointwise CNN for feature enhancement, and Spatial and Temporal Patching modules for organizing features into spatial-temporal patches, EEG-PatchFormer jointly learns spatial-temporal information from EEG data. Leveraging the global learning capabilities of the self-attention mechanism, it captures essential features across brain regions over time, thereby enhancing EEG data decoding performance. Demonstrating superior performance, EEG-PatchFormer surpasses existing benchmarks in accuracy, area under the ROC curve (AUC), and macro-F1 score on a public cognitive attention dataset. The code can be found via: https://github.com/yi-ding-cs/EEG-PatchFormer .
\end{abstract}

\section{Introduction}
Originally developed for enabling communication and control for paralyzed patients, Brain-Computer Interface (BCI) systems have expanded their applications to include cognitive training and assessments, modeling and evaluating the cognitive states of healthy individuals \cite{Abiri_2019}. In EEG-based BCI experiments, a small number of electrodes are positioned on the subject's head, making it a cost-effective, non-invasive method. This accessibility has made EEG a popular choice in passive BCI research areas and studies focusing on attention \cite{7902094,Lim2012}. Attention is crucial for numerous daily activities requiring focus, such as studying and operating machinery \cite{https://doi.org/10.1111/j.1365-2869.2010.00877.x}. EEG-based BCI systems have been effective in enhancing attention, showing significant improvements in ADHD symptoms \cite{Lim2012}.

With the rapid development of deep learning in domains such as computer vision \cite{he2016deep,dosovitskiy2021an, qin2023data, Li_2019_CVPR, Li_2023_CVPR, Pan_Gao_Zhang_Zheng_Shen_Li_Hu_Liu_Dai_2024,li2024adaptive,liintegrating}, natural language processing (NLP) \cite{kenton2019bert, ding-etal-2023-gpt}, multimodal learning\cite{gao2024cantor}, and graphs \cite{kipf2017semisupervised,10366850}, deep learning methods catch more and more attention from researchers in the BCI field and outperform traditional machine learning approaches in achieving higher accuracy for EEG classification tasks \cite{8634938,9206750,9762054,doi:10.1002/hbm.23730,doi:10.1080/27706710.2023.2181102}. For instance, Lawhern et al. \cite{Lawhern_2018} introduced EEGNet, which excels in processing both spatial and temporal data using convolutional layers. Furthermore, TSception \cite{9206750, 9762054} utilized multi-scale temporal and spatial CNN layers to extract temporal dynamics and the spatial asymmetric patterns. Ding et al. \cite{10025569} drew inspiration from neurophysiological knowledge and designed a local-global graph neural network, named LGGNet, which achieves promising EEG decoding performance across a variety of cognitive tasks, including attention, fatigue, emotion, and preference. However, these methods learn spatial-temporal information in separate layers, which might not effectively extract complex spatial-temporal information.

To jointly learn spatial-temporal information, we introduce the EEG-PatchFormer, a novel approach for decoding attention states from EEG signals. This innovative architecture comprises five essential components: a Temporal CNN module for initial EEG feature extraction across frequencies; a Feature Enhancement module to refine these features; Spatial and Temporal Patching modules that segment the EEG data into structured patches capturing both local and global spatial-temporal dynamics; and a Transformer module that delves into the interconnections among these patches, elucidating complex spatial-temporal relationships. The intricate processing pipeline culminates in a fully connected layer that classifies the attention states, employing a sophisticated transformer encoder and attention mechanism to distill and interpret the salient features across different brain regions and temporal spans. This methodological advancement offers a nuanced perspective on EEG data, promising enhanced decoding accuracy. When benchmarked against several established methods on a public attention dataset, EEG-PatchFormer demonstrates superior performance, showcasing its potential to significantly enhance attention state decoding.
 
The major contribution of this work can be summarised as: 
\begin{itemize}
\item We introduce a novel spatial patching module featuring local and global branches for generating spatial EEG patches, complemented by a temporal patching module for creating overlapped spatial-temporal EEG patches.
\item We present EEG-PatchFormer, a Transformer-based approach for decoding attention states from spatial-temporal patches.
\item EEG-PatchFormer outperforms multiple baseline methods in attention state classification on a public benchmark dataset, achieving superior classification results. 
\end{itemize}

\begin{figure}[tp]
    \centering
    \includegraphics[width=\linewidth]{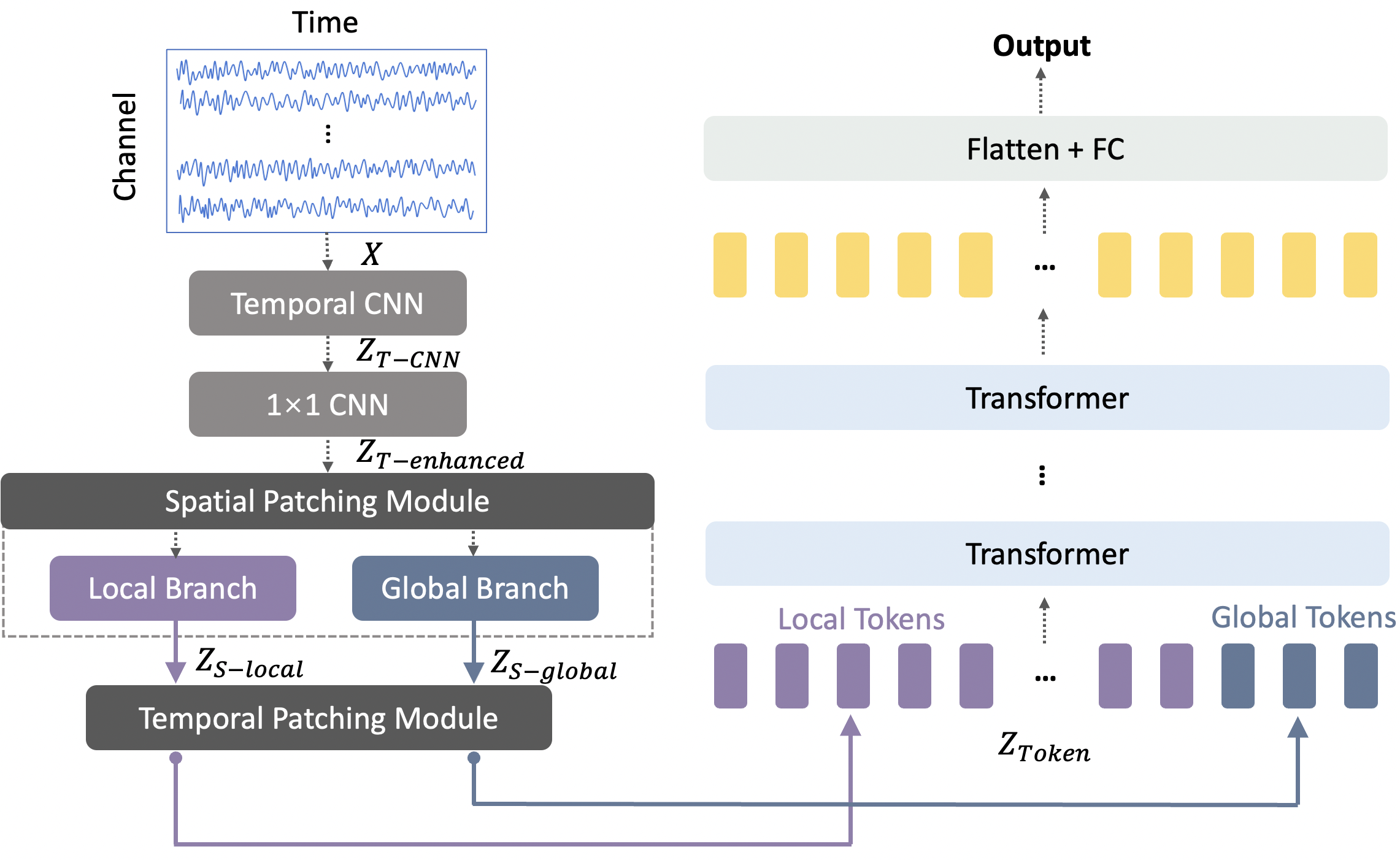}
    \caption{Neural network architecture of EEG-PatchFormer. It begins with a Temporal CNN module for initial feature extraction across frequencies, followed by a Feature Enhancement module to refine these features. Spatial and Temporal Patching modules then segment the data into structured patches to capture local and global spatial-temporal dynamics. A Transformer module \cite{NIPS2017_3f5ee243} analyzes the interconnections among these patches, revealing complex spatial-temporal relationships. Finally, a fully connected layer, utilizing a sophisticated transformer encoder and attention mechanism, classifies attention states by distilling and interpreting salient features across various brain regions and temporal spans.}
    \label{fig:patchformer}
\end{figure}

\section{Method}
\subsection{Temporal CNN Module}
To learn the dynamic temporal information within different frequency bands, a 1-dimensional convolutional layer along the temporal dimension, known as the temporal learner, is applied to the EEG data. This layer is designed to extract temporal features, capturing informative patterns within the time series. It comprises a 1D-CNN layer, leaky ReLU activation, a batch normalization layer, and an average pooling layer. Let's denote an EEG input and the corresponding output layer by $X \in \mathbb{R}^{c \times l}$ and $y\in \mathbb{R}$, where $c$ is the number of EEG channels, and $l$ is the length of the input EEG sample. This module can be formalised as follows
\begin{equation}
    Z_{T-CNN}=\textrm{AvgPool}(\Phi_{\textrm{LeakReLU}} (\textrm{BN}(\textrm{CNN}(X)))
\end{equation}
where $\Phi_{\textrm{LeakReLU}}(\cdot)$ is the LeakyReLU activation function, $\textrm{BN}(\cdot)$ is the batch normalization layer, $\textrm{AvgPool}(\cdot)$ is the average pooling layer. The kernel size of the CNN layer is set to $\frac{f_s}{2}$, where $f_s$ is the sampling rate of the EEG. The number of the CNN kernels in this layer is set as $k=32$. The same padding technology is utilized in CNN layers. As the pooling length and step of the average pooling are both 4, the output has a shape of $(k \times c \times \frac{l}{4})$.

\subsection{Feature Enhancement Module}
One by one convolutions, also known as pointwise convolutions, can enhance the learned features from different CNN kernels \cite{10025569}. The output, $Z_{T-CNN} \in \mathbb{R}^{k \times c \times \frac{l}{4}}$ is then passed through a 1x1 convolution to attentively enhance the features and effectively reduce the dimensionality of the data. The output, $Z_{T-enhanced} \in \mathbb{R}^{k \times c \times \frac{l}{8}}$ can be calculated by
\begin{equation}
    Z_{T-enhanced}=\textrm{AvgPool}(\Phi_{\textrm{LeakReLU}}(\textrm{BN}(\textrm{CNN}(X)))).
\end{equation}
The pooling size and step are both 2, and the number of the CNN kernels is the same as the one in temporal CNN layer as suggested in LGGNet \cite{10025569}.

\subsection{Spatial Patching Module}
\begin{figure}[tp]
    \centering
    \includegraphics[width=\linewidth]{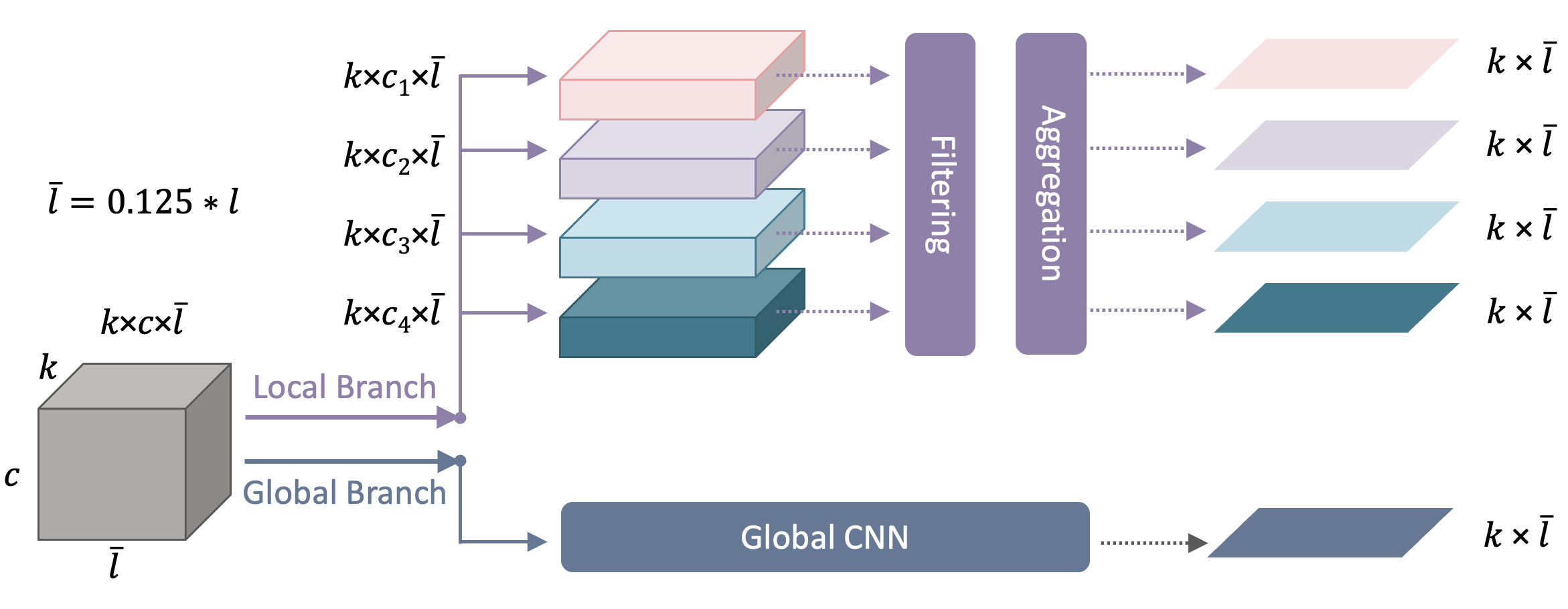}
    \caption{Diagram of the spatial patching module. There are two branches in SPM: local branch and global branch. The $c_n$ represent the number of channels within each local graph. It may be different for different local graphs depends on the deviation of local graphs. We use general graph, $G_{general}$, defined in LGGNet, as the deviation of local graphs in our study. An input of size $(k \times c \times \tilde{l})$ is used here to illustrate the mechanism of SPM.}
    \label{fig:SPM}
\end{figure}

There are two branches in the Spatial Patching Module (SPM): 1) the local branch and 2) the global branch. The data is passed into these two parallel branches, one tasked with extracting global spatial information and the other with learning local information, resulting in both global and local spatial patches. Fig.~\ref{fig:SPM} depicts the architecture of the spatial patching module.

\begin{table}[t]
\centering
\ra{0.6}
\caption{The detailed configuration of the local graphs.}
\begin{tabularx}{0.8\linewidth}{
  @{\extracolsep{\fill}\hspace{\tabcolsep}}
  c c c
} 
\toprule
Index $n$ & $c_{n}$ & Channels\\
\midrule 
1 & 2 & Fp1, Fp2\\
2 & 3 & AFF5, AFz, AFF6\\
3 & 2 & F1, F2\\
4 & 4 & FC5, FC1, FC2, FC6\\
5 & 3 & C3, Cz, C4\\
6 & 4 & CP5, CP1, CP2, CP6\\
7 & 5 & P7, P3, Pz, P4, P8\\
8 & 1 & POz\\
9 & 2 & O1, O2\\
10 & 1 & T7\\
11 & 1 & T8\\
\bottomrule
\end{tabularx}
\begin{tablenotes}
      \small
      \item  $c_{n}$: the number of EEG channels within $n$-th local graph.
    \end{tablenotes}
\label{Tab:local_graphs}
\end{table}

\subsubsection{Local spatial learner} The local branch focuses on capturing features specific to individual brain regions. Following \cite{10025569}, we group the EEG channels into several local groups as shown in TABLE~\ref{Tab:local_graphs}, and perform local graph filtering on the grouped EEG representations. The input is first reshaped using the rearrange function to prepare it for local processing, resulting a tensor, $Z_{reshaped} \in \mathbb{R} ^{c \times \frac{k*l}{8}}$. A learnable local filter with a learnable weight matrix $W_{local}$ which has the same size as $Z_{reshaped}$, is then applied element-wise to emphasize potentially informative features from different brain areas. This filter is combined with a bias term, $b_{local}$, to account for baseline variations as used in \cite{10025569}. The filtered representation, $Z_{filtered}\in \mathbb{R} ^{c \times \frac{k*l}{8}}$, can be calculated by

\begin{equation}
  Z_{filtered}=\Phi_{ReLU}(W_{local}\circ Z_{reshaped}-b_{local})
\end{equation}
where $\Phi_{ReLU}(\cdot)$ is the ReLU activation function, and $\circ$ is the Hadamard product.

Subsequently, an aggregator module combines these filtered features from various brain regions by taking the mean of their values. This aggregation step produces a consolidated feature vector representing each brain region. The output of the local branch, $Z_{S-local}\in \mathbb{R} ^{1 \times \frac{k*l}{8}}$, can be calculated by

\begin{equation}
  Z_{S-local}=\textrm{Aggregate}(Z_{filtered}, \mathcal{R}_{local}) 
\end{equation}
where $\mathcal{R}_{local}$ is a list that defines the EEG channels within each local brain region. 

The $Aggregate(\cdot)$ can be described as

\begin{algorithm}
  \caption{Aggregation on local brain areas}\label{euclid}
  \begin{algorithmic}[1]
    \Procedure{Aggregate}{$Z_{filtered}, \mathcal{R}_{local}$}
      \For{\texttt{$c_{local}^{i}$ in $\mathcal{R}_{local}$}}
        \State \texttt{$num_{c}^{i} = \textrm{len}(c_{local}^{i})$}
        \For{\texttt{j in $c_{local}^{i}$}}
         \State \texttt{$Z_{S-local}^{i} = \frac{1}{num_{c}^{i}}\sum_{j=1}^{num_{c}^{i}}Z_{filtered}^{j,:}$}
      \EndFor
      \State \Return $Z_{S-local}^{i}$\Comment{Vector of one local region.}
      \EndFor
      \State \Return $[Z_{S-local}^{i}],i=1,2,..., c_{n} $\Comment{Stack the vectors of all local regions.}
    \EndProcedure
  \end{algorithmic}
\end{algorithm}

\subsubsection{Global spatial learner} The global spatial learner aims to extract overarching characteristics across all EEG channels. A final convolution layer, known as the global CNN, aggregates these features, resulting in a single global feature vector that encapsulates information from all channels across the entire time window. The output of the global branch, $Z_{S-global}$, can be calculated by
\begin{equation}
  Z_{S-global}=\textrm{AvgPool}(\Phi_{\textrm{leakReLU}}(\textrm{BN}(\textrm{CNN}(X)))) 
\end{equation}
where the number of the CNN kernel is set as $k=32$ as well, and the kernel size is $(c,1)$. At the end of both spatial learning branches, the resulting global and local feature vectors are concatenated to get the final output of SPM, $Z_S \in \mathbb{R}^{p \times k \times \frac{l}{8}}$, where $p$ is the number of the spatial patches
\begin{equation}
  Z_S=[Z_{S-local},Z_{S-global}]
\end{equation}
where $[\cdot]$ represents the concatenation operation.

\subsection{Temporal Patching Module}
\begin{figure}[tp]
    \centering
    \includegraphics[width=\linewidth]{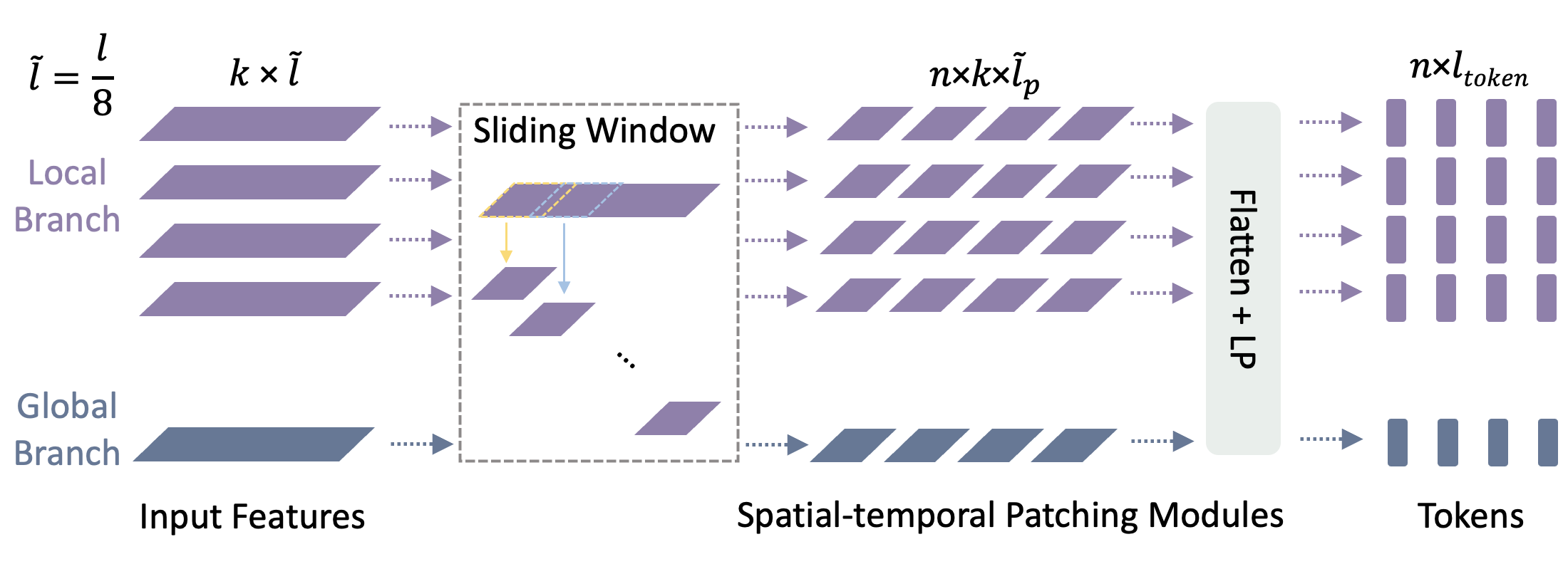}
    \caption{The pipeline of the temporal patching module. }
    \label{fig:TPM}
\end{figure}

To capture the temporal dynamics of the EEG, a temporal patching module (TPM) is utilized to split the learned spatial patches into $q$ overlapped temporal patches by a sliding window with a length of $l_t$ and a step of $l_{step}$.  The pipeline of TPM is shown in Fig.~\ref{fig:TPM}. After processing with the sliding window, the patches are flattened and linearly projected by a Linear Projection layer (LP) with the learnable weight $W_{LP}$ and bias $b_{LP}$. Consequently, the final spatial-temporal tokens from TPM, $Z_{token} \in \mathbb{R}^{q \times l_{token}}$, can be calculated by
\begin{equation}
    Z_{token}=LP(\Gamma_{flatten}(\Gamma_{window} (Z_S)))
\end{equation}
where  $\Gamma_{window}(\cdot)$ is the sliding window function, and $\Gamma_{flatten}(\cdot)$ is the flatten operation.

\subsection{Transformer Module}
Given the spatial-temporal tokens, $Z_{token} \in \mathbb{R}^{q \times l_{token}}$, a transformer module \cite{NIPS2017_3f5ee243} with 4 transformer layers is applied to capture the spatiotemporal information. 

Finally, the output from the Transformer module is flattened and fed into a fully connected layer to get the classification output. 

\section{Experiment}
\subsection{Dataset}
We assessed the proposed EEG-PatchFormer model using a publicly accessible EEG dataset from Shin et al. \cite{Shin2018}, specifically designed for cognitive attention classification tasks (high and low attention). This comprehensive dataset comprises 28-channel EEG recordings from 26 participants engaged in Discrimination/Selection Response (DSR) tasks, focusing on cognitive load variations. Each experimental session was structured with multiple cycles of 40-second active task periods followed by 20-second rest intervals, providing a balanced mix of cognitive engagement and recovery phases. This setup allowed us to rigorously evaluate our model's capability to decode and classify cognitive states from EEG signals under varied attentional demands.
\subsection{Pre-processing}
Adhering to the preprocessing steps outlined by Ding et al. \cite{10025569}, we first employed independent component analysis (ICA) to eliminate artifacts from the EEG data. Subsequently, the data was downsampled to 250 Hz to manage computational load while retaining signal integrity. To ensure class balance, only the initial 20 seconds of each task were utilized, with trials being segmented through a 4-second sliding window approach, incorporating a 50\% overlap between consecutive segments. This methodology facilitated a detailed and balanced analysis of cognitive states within the EEG dataset.

\subsection{Experiment Settings}
We adopted a leave-one-subject-out (LOSO) cross-validation to evaluate the proposed method. Each subject’s data was isolated for testing, and the remaining data was further partitioned into 80\% for training and 20\% for validation. This approach not only aids in mitigating overfitting but also enhances the model's generalisability, affirming its robustness and reliability across diverse individual profiles. Additionally, this rigorous validation framework supports a thorough assessment of the model’s ability to adapt and perform consistently, irrespective of the variability inherent in individual EEG patterns.
\subsection{Evaluation Metrics}
We employed three standard evaluation metrics to assess EEG-PatchFormer’s performance in subsequent experiments:
1) Mean accuracy (ACC): The percentage of correctly classified attention states across all test samples.
2) Area under ROC curve (AUC): The Receiver Operating Characteristic (ROC) curve depicts the trade-off between true positive rate and false positive rate for various classification thresholds. AUC summarises the overall performance of a model across all possible thresholds by calculating the area under the ROC curve. A higher AUC value indicates better classification ability.
3) Macro-F1 Score:  The F1 score is a harmonic mean between precision and recall in multi-class classification problems. Macro-F1 averages F1 scores across all individual classes, presenting a balanced measure of the model’s accuracy across all categories.

\subsection{Implementation Details}

Cross-entropy loss directs the training process, with the Adam optimizer set to an initial learning rate of 1e-3 and weight decay at 1e-5. We employ a cosine annealing schedule for learning rate adjustment during training. To prevent overfitting, dropout rates of 0.5 are applied. We use a batch size of 64. The model undergoes training for 200 epochs, selecting the iteration with the highest validation accuracy for test data evaluation. CNN layer kernel lengths are determined by $0.5*f_{s}$, with $f_{s}$ representing the EEG segment's sampling rate, resulting in a kernel size of (1, 100). We configure the CNN with 32 kernels. The temporal patching module (TPM) utilizes a sliding window function with a length of 20 and a step size of 5. The transformer architecture comprises 4 layers, and we set the number of attention heads and the dimension of the attention mechanism, $n_{head}$ and $d_{atten}$, to 32 each.
\begin{table}[t]
\centering
\ra{0.6}
\caption{Performance comparison for attention detection}
\begin{tabularx}{\linewidth}{
  @{\extracolsep{\fill}\hspace{\tabcolsep}}
  c c c c
} 
\toprule
Methods & ACC (\%) & AUC & F1-macro (\%) \\
\midrule 
DGCNN & 60.98$\pm$6.05	&0.70$\pm$0.07	&56.81$\pm$9.51\\ 
LGGNet &67.81$\pm$6.38	&0.76$\pm$0.08	&66.70$\pm$7.69\\
TSception &71.60$\pm$7.66	&0.82$\pm$0.07	&69.94$\pm$10.88\\
ViT	&67.72$\pm$7.03	&0.75$\pm$0.08	&66.79$\pm$8.49\\
EEG-PatchFormer	&\textbf{75.63}$\pm$7.33	&\textbf{0.85}$\pm$0.07	&\textbf{75.04}$\pm$8.03\\ \bottomrule
\end{tabularx}
\label{Tab:result_loso}
\end{table}

\begin{table}[t]
\centering
\ra{0.6}
\caption{Ablation study results}
\begin{tabularx}{\linewidth}{
  @{\extracolsep{\fill}\hspace{\tabcolsep}}
  c c c c c c
} 
\toprule
&ACC (\%) & AUC & F1-macro (\%) \\
\midrule 
w/o FEM &71.36$\pm$6.90	&0.81$\pm$0.07	&70.47$\pm$7.70 \\
w/o SPM&64.62$\pm$7.76	&0.72$\pm$0.09	&63.37$\pm$8.93 \\
w/o OTPM&73.99$\pm$6.95	&0.83$\pm$0.07	&73.25$\pm$7.95 \\
EEG-PatchFormer&\textbf{75.63}$\pm$7.33	&\textbf{0.85}$\pm$0.07	&\textbf{75.04}$\pm$8.03\\
\bottomrule
\end{tabularx}
\begin{tablenotes}
      \small
      \item FEM: Feature enhancement module. SPM: Spatial patching module. OTPM: Overlap in Temporal patching module.
    \end{tablenotes}
\label{Tab:result_ablation}

\end{table}

\begin{table}[t]
\centering
\ra{0.6}
\caption{Effects of the patch lengths in TPM}
\begin{tabularx}{\linewidth}{
  @{\extracolsep{\fill}\hspace{\tabcolsep}}
  c c c c
} 
\toprule
Patch length & ACC (\%)& AUC & F1-macro (\%) \\
\midrule 
10 & 75.17$\pm$7.76	&0.84$\pm$0.07	&74.42$\pm$8.96\\ 
20 &\textbf{75.63}$\pm$7.33	&\textbf{0.85}$\pm$0.07&\textbf{75.04}$\pm$8.03\\
30 &74.47$\pm$7.57	&0.84$\pm$0.07	&73.63$\pm$8.61\\
40	&74.86$\pm$7.07	&0.84$\pm$0.07	&74.15$\pm$8.07\\
50	&75.55$\pm$6.95	&0.85$\pm$0.07	&74.88$\pm$7.80\\
\bottomrule
\end{tabularx}
\label{Tab:result_patch}

\end{table}

\section{Results and analyses}

The main experiment involved benchmarking EEG-PatchFormer’s accuracy against established baseline models. We compared its performance on the target dataset against that of DGCNN \cite{8320798}, LGGNet \cite{10025569}, TSception \cite{9762054}, and ViT \cite{dosovitskiy2021an}, seeking to gauge its effectiveness and determine if it offers improvements in classification accuracy. TABLE~\ref{Tab:result_loso} displays the mean accuracy, AUC, and macro-F1 score of EEG-PatchFormer in comparison to those of the baseline models.

Out of all the models tested, EEG-PatchFormer achieved the highest classification accuracy. Specifically, the mean accuracy achieved by EEG-PatchFormer was 75.63\%, a 4.03\% increase from the next highest performing model, TSception. Likewise, EEG-PatchFormer had the highest AUC, a value of 0.85 which once again outperformed the next best model, TSception, by 0.03. Finally, EEG-PatchFormer also achieved the highest macro-F1 score of 75.04\%. This was 5.1\% higher than TSception in second.

Overall, our proposed model significantly outperforms the established baseline models across all three performance metrics. This justifies our hypothesis that optimal EEG attention classification performance may be accomplished by a transformer-based network that meaningfully captures spatiotemporal dynamics inherent in EEG signals.

\subsection{Ablation Study}
We conducted ablation studies to analyze the impact of EEG-PatchFormer’s components: (1) Overlapping vs. non-overlapping patches – Non-overlapping patches reduced accuracy from 75.63\% to 73.99\%, highlighting the importance of feature continuity. (2) Feature enhancement module – Removing the 1x1 convolution layer dropped accuracy to 71.36\%, demonstrating its role in feature extraction and dimensionality reduction. (3) Spatial patching module – Eliminating this module caused a sharp accuracy decline to 64.62\%, underscoring its necessity for spatial information integration. Table~\ref{Tab:result_ablation} summarizes these findings, reinforcing the significance of each component in EEG-PatchFormer’s performance.

\subsection{Effect of The Patch Lengths in TPM}
To better understand the effect of the patch lengths of TPM, we conducted some experiments in which the patch length is increased from 10 to 50 with a step of 10. Increasing the patch lengths can increase the temporal information contained in each patch. The results are shown in TABLE ~\ref{Tab:result_patch}. The optimal patch length for our proposed model was found to be 20 data points, producing the highest scores for all three metrics. Specifically, this configuration achieved a mean validation accuracy of 75.63\% (0.08\% higher than the next best configuration), AUC of 0.85, and macro-F1 of 75.04\% (0.16\% higher).

\section{Conclusion}
In this paper, we proposed EEG-PatchFormer, a novel transformer-based deep learning model tailored for EEG attention classification in BCI applications. EEG-PatchFormer features five main modules, including a temporal CNN, feature enhancement module, spatial and temporal pathcing modules, and a transformer, significantly outperforming benchmarks in accuracy, AUC, and macro-F1 score on a cognitive attention dataset. Despite its success, the model's evaluation on a single dataset calls for broader validation to confirm its generalizability in the BCI field. Future directions involve comprehensive hyperparameter tuning and ablation studies to enhance performance further. Additionally, exploring EEG-PatchFormer's applicability to other EEG-based cognitive state classifications, like fatigue and emotion, presents an exciting avenue for expansion.


\section*{Acknowledgment}
This work was supported by the RIE2020 AME Programmatic Fund, Singapore (No. A20G8b0102) and the MoE AcRF Tier 1 Project (No. RT01/21).

\bibliographystyle{IEEEtran}
\bibliography{mybib}
\vspace{12pt}

\end{document}